\documentclass[fleqn,usenatbib]{mnras}

\usepackage{newtxtext,newtxmath}

\usepackage[T1]{fontenc}

\DeclareRobustCommand{\VAN}[3]{#2}
\let\VANthebibliography\thebibliography
\def\thebibliography{\DeclareRobustCommand{\VAN}[3]{##3}\VANthebibliography}

\usepackage{graphicx}	
\usepackage{amsmath}	
\usepackage{tikz,xcolor,hyperref} 

\definecolor{lime}{HTML}{A6CE39}
\DeclareRobustCommand{\orcidicon}{%
	\begin{tikzpicture}
	\draw[lime, fill=lime] (0,0) 
	circle [radius=0.16] 
	node[white] {{\fontfamily{qag}\selectfont \tiny ID}};
	\draw[white, fill=white] (-0.0625,0.095) 
	circle [radius=0.007];
	\end{tikzpicture}
	\hspace{-2mm}
}

\foreach \x in {A, ..., Z}{%
	\expandafter\xdef\csname orcid\x\endcsname{\noexpand\href{https://orcid.org/\csname orcidauthor\x\endcsname}{\noexpand\orcidicon}}
}

\title[Binary Properties of Rotational Variables]{Statistical Estimates of the Binary Properties of Rotational Variables}

\author[A. Phillips et al.]{
Anya Phillips\orcidA{}$^{1}$\thanks{E-mail: phillips.1671@osu.edu} and 
C. S. Kochanek$^{1,2}$
\\
$^{1}$Department of Astronomy, The Ohio State University, 140 West 18th Avenue, Columbus, OH, 43210, USA\\
$^{2}$Center for Cosmology and Astroparticle Physics, The Ohio State University, 191 W. Woodruff Avenue, Columbus, OH, 43210, USA
}

\date{Accepted XXX. Received YYY; in original form ZZZ}

\pubyear{\the\year{}}

\begin{document}
\label{firstpage}
\pagerange{\pageref{firstpage}--\pageref{lastpage}}
\maketitle

\begin{abstract}
We present a model to estimate the average primary masses, companion mass ranges, the inclination limit for recognizing a rotational variable, and the primary mass spreads for populations of binary stars. The model fits a population's binary mass function distribution and allows for a probability that some mass functions are incorrectly estimated. Using tests with synthetic data, we assess the model's sensitivity to each parameter, finding that we are most sensitive to the average primary mass and the minimum companion mass, with less sensitivity to the inclination limit and little to no sensitivity to the primary mass spread. We apply the model to five populations of binary spotted rotational variables identified in ASAS-SN, computing their binary mass functions using RV data from APOGEE. Their average primary mass estimates are consistent with our expectations based on their CMD locations ($\sim0.75 M_{\odot}$ for lower main sequence primaries and $\sim 0.9$--$1.2 M_{\odot}$ for RS CVn and sub-subgiants). Their companion mass range estimates allow companion masses down to $M_2/M_1\simeq0.1$, although the main sequence population may have a higher minimum mass fraction ($\sim0.4$). We see weak evidence of an inclination limit $\gtrsim50^{\circ}$ for the main sequence and sub-subgiant groups and no evidence of an inclination limit in the other groups. 
No groups show strong evidence for a preferred primary mass spread. We conclude by demonstrating that the approach will provide significantly better estimates of the primary mass and the minimum mass ratio and reasonable sensitivity to the inclination limit with 10 times as many systems.

\end{abstract}




\section{Introduction}

Population-level properties of binary stars are important to understanding their formation and evolution. 
At high masses (primaries $\gtrsim 3 M_{\odot}$; \citealt{moe2017}), these systems can evolve to produce X-ray binaries \citep{verbunt1993}, pulsars \cite{Lorimer2008}, and Type Ia supernovae \citep{wang2012}, among other extreme astrophysical phenomena. %
Across stellar mass, however, measuring the fundamental properties of binary systems provides insight into binary star formation. An example is the distribution of binary mass ratios. Observing companion masses distributed as the stellar initial mass function (IMF) implies that each companion has evolved independently of its primary \citep{tout1991,mcdonald1995}. On the other hand, correlated companion and primary masses imply that the two stars coevolved through pre-main sequence fragmentation, fission, or mass transfer through Roche Lobe overflow \citep{bonnell1992,kroups1995_1,kroups1995_2,Clarke1996,kratter2006,kouwenhoven2009,marks2011,bate2012,moe2017}. Another point of interest is the relation between multiplicity or close binary fraction on stellar mass (e.g., \citealt{duquennoy91, Duchene2013, mazzola2020}).

Measuring binary properties also provides an observational foundation for binary star evolution. Numerous studies have, for example, examined the physics of tidal circularization  \citep{koch81, duquennoy91, verbuntphinney1995, meibom05, vaneylen16, bashi23}, tidal synchronization (e.g. \citealt{lurie2017, Leiner2022}), and their dependence on stellar mass, binary mass ratio, eccentricity, and orbital period. Finally, measured binary properties inform binary population synthesis models, as the rates of different channels of binary evolution depend on the assumed binary statistics (see \citealt{han2020} for a recent review).

Unresolved main sequence binaries can be identified photometrically, sitting above the main sequence in luminosity, especially for unevolved primaries ($T_{\rm{eff}}<5250\ K$). \citet{simonian2019} use main sequence rotational variables in the \textit{Kepler} to show that rapid rotation (rotation periods $<7$ days) occurs primarily in photometric binaries. For these systems, the apparent elevation above the main sequence allows an estimate of the binary mass ratio, but this method is best for detecting near equal mass binaries because of the steepness of the mass-luminosity relation on the main sequence.

Eclipsing double lined spectroscopic binaries (SB2s) provide the most information about a system's fundamental binary properties, allowing measurement of both companion masses and the orbit's inclination (e.g., \citealt{torres2010}). Non-eclipsing SB2s allow measurement of the mass ratios of individual systems, but their inclinations, and hence, also their masses, are still unknown. 
Identifying SB2s is straightforward at periods $\lesssim 5$ years, but at any orbital period it is also possible to spectroscopically disentangle binaries by fitting a binary spectral model rather than a single-star model, recovering the mass ratio and atmospheric composition of both companions (e.g., \citealt{el-badry2018}).
Binary samples with spectroscopic signals from both stars are also biased toward equal mass companions because high mass ratio systems typically have a high flux ratio and a secondary that is difficult to detect.

Eclipsing single lined spectroscopic binaries (SB1s) require knowledge of the primary mass in order to recover the secondary mass. For non-eclipsing SB1s, we can only observe the orbital period $P$ and velocity semiamplitude $K$ to compute the binary mass function,
\begin{equation}
    \label{eq:f(M)}
    f = \frac{P K^3}{2\pi G} = \frac{M_2^3\sin^3{i}}{(M_1 + M_2)^2} = \frac{q^3 M_1 \sin^3{i}}{(1+q)^2},
\end{equation}
where $q$ is the ratio of the secondary to primary mass $M_2/M_1$, and we have assumed a circular orbit ($e=0$). However, the inclination is still unknown, so we can only estimate the binary mass ratio as a function of inclination, even with an estimate of the primary mass.

Where the primary masses are known, there are multiple statistical approaches to deriving the distribution of binary mass ratios despite the problem of unknown inclinations. One approach is using an averaged value of $\sin^3 i$ (e.g., \citealt{aitken1964}), but this is valid only for very large samples of binaries. Other approaches involve assuming a mass ratio distribution and comparing the implied distribution in an observable like the binary mass function to observations, either assuming randomly-oriented orbits \citep{jaschek1972,halbwachs1987,trimble1990}, or iteratively adjusting the assumptions about the inclination and mass ratio distributions \citep{MazehGoldberg1992}. 
The underlying distribution of inclinations is also of interest, both for binary orbits and single rotating stars. For example, \citet{jackson2010} model the $\sin i$ distributions for G- and K-type stars in the Pleiades and Alpha Persei clusters and find that rotational modulation from starspots can be detected down to inclinations of $30^{\circ}$.

For systems with main sequence primaries, color (or luminosity) is a good proxy for primary mass, allowing use of these types of methods for SB1 orbits. Particularly within star clusters, it is also possible to estimate the mass of a main sequence primary with a stellar evolutionary track or isochrone. In these cases, the binary mass ratio is of central interest. For example, \citet{geller2021} convert distributions in primary mass and binary mass function to distributions in secondary mass and mass ratio for main sequence binaries in M67 using the algorithm from \citet{MazehGoldberg1992}. They find a uniform distribution of mass ratios ranging from $\sim 0.1$--$1$, rather than IMF-distributed companion masses. 
\citet{Raghavan2010} also find a roughly uniform distribution of mass ratios for main sequence field binaries from \textit{Hipparchos}, but with an overabundance of ``twin" systems (with mass ratios $q\geq0.95$). 
\citet{pinsonneault2006} show that 50\% of a sample of OB-type detached eclipsing binaries in the Large Magellanic Cloud are twin systems with $q>0.87$.
For close binaries, uniformly distributed mass ratios, or correlated companion masses, are expected and usually observed (e.g., \citealt{tokovinin2000, sana2011}).

Estimating stellar mass based on luminosity is difficult for evolved primaries, since there is no mass-luminosity relation for giants. It is therefore more difficult to measure the binary properties of SB1s with red giant primaries. \citet{Leiner2022} used the distribution of their sample of heavily spotted red giant binaries (RS CVn; \citealt{hall1976}) in luminosity versus rotation period to provide a rough upper limit on their primary masses of $1$--$2 M_{\odot}$. They assumed edge-on, circular, and tidally locked orbits, and matched the maximum radius the primary could have before exceeding its Roche lobe to the corresponding stellar mass on an isochrone. They also examined a sample of sub-subgiants, an under-luminous and shorter-period subclass of RS CVn, which tend to have lower masses and tidally synchronize only at shorter periods ($\lesssim10$ days) than the typical period range for RS CVn ($\sim$tens of days). 

Here we take a slightly different statistical approach to previous work, presenting a method to estimate the average primary mass, minimum mass ratio, minimum inclination limit for recognizing rotational variables, and primary mass spread for binaries at the population level. The method does not require full SB2 statistics, eclipses, or a prior estimate of the primary mass. 
In doing so, we aim to understand the five groups of binary spotted rotational variable stars in the All-Sky Automated Survey for Supernovae (ASAS-SN) variable star catalogue \citep{jayasinghe2018,jayasinghe2019II,jayasinghe2019I,jayasinghe2020, jayasinghe2021, christy23} examined in \citet{Phillips2024}. We fit circular orbits at the ASAS-SN rotation period to sparse radial velocity (RV) curves from the Apache Point Observatory Galactic Evolution Experiment (APOGEE; \citealt{Majewski2017}) to determine their binary mass functions. We then fit each group's binary mass function distribution using Monte Carlo simulations to estimate the parameters. 
In \citet{Phillips2024}, we found that each group consists of tidally synchronized or pseudosynchronized systems.
The MS2 stars are binary main sequence stars with periods of $\lesssim 10$ days. 
The G1 and G3 stars are RS CVn binaries with periods of tens of days.
The G2 stars are RS CVn and sub-subgiant binaries, with lower luminosity primaries and shorter periods ($\sim$10 days) than G1/3. 
The G4 stars have intermediate luminosities to G1-3 but with rotation periods approaching 100 days that tend to be modestly longer than their binary orbital periods. 
We ignore the MS1 group since it consists of single stars.

In Section \ref{sec:2} we describe the data used, our method for fitting circular orbits to determine each system's mass function, and the Monte Carlo simulations we use to fit the mass function distributions. We present the resulting fits and discuss implications for each population of binaries in Section \ref{sec:3}, concluding and discussing future directions in Section \ref{sec:4}.

\section{Methods}\label{sec:2}

We use RVs from the \texttt{allVisit} catalog of APOGEE DR17 \citep{apogeedr172022}. We require a signal to noise ratio of $>$20, that \texttt{N\_COMPONENTS}$\neq$0, and that there are $\geq4$ RV measurements. With $\geq4$ RVs we can fit a circular orbit with at least one degree of freedom as a check of the fit.
We assume the binaries have circular orbits with the photometric period of the rotational variable (see \citealt{lurie2017} for a discussion of tidal synchronization in orbits with periods $\lesssim$ tens of days). This appeared to be true for all but the G4 group, which tended to be slightly subsynchronous.
The RV curves are fit as
\begin{equation}
    \label{eq:circular_orbit}
    v = v_0 + a\cos(2\pi(t-t_0)/P) + b\sin(2\pi(t-t_0)/P)
\end{equation}
for $v_0$, $a$, and $b$. The velocity semiamplitude $K = \sqrt{a^2 + b^2}$ and the orbital period $P$ provide the system's binary mass function (Equation \ref{eq:f(M)}).

Because the time span used to determine periods from the ASAS-SN light curve ($\Delta T_A \simeq 5$~years) is shorter than the time span of some of the APOGEE data ($\Delta T_s \simeq 9$~years) and because the precision of rotational periods is limited by star spot evolution, we allow small shifts in the period when fitting the APOGEE velocities. A period shift of $\Delta P$ leads to a phase shift of $\Delta n=\Delta T\Delta P/P^2$ over the period $\Delta T$, so the period accuracy needed for APOGEE is $\Delta P\simeq0.1 P^2/\Delta T$ (allowing a 10\% phase shift over the span of the data).
We fit circular orbits over the period range $P_{\rm{rot}}\pm 30 P^2/\Delta T_A$ in steps of $\frac{\Delta P}{10} \frac{\Delta T_S}{\Delta T_A}$. 
From this best fit, we perform a refined local search over $P_{\rm{new}}\pm P^2_{\rm{new}}/\Delta T_A$ in steps of $\Delta P_{\rm{new}}/100$.
\citet{Phillips2024} found that the reported rotation period was sometimes half the true period for the G1 group of spotted giants, so for these stars, we also carried out fits assuming $P_{\rm{orb}}=2P_{\rm{rot}}$.

\begin{figure}
    \centering
    \includegraphics[width=0.85\columnwidth]{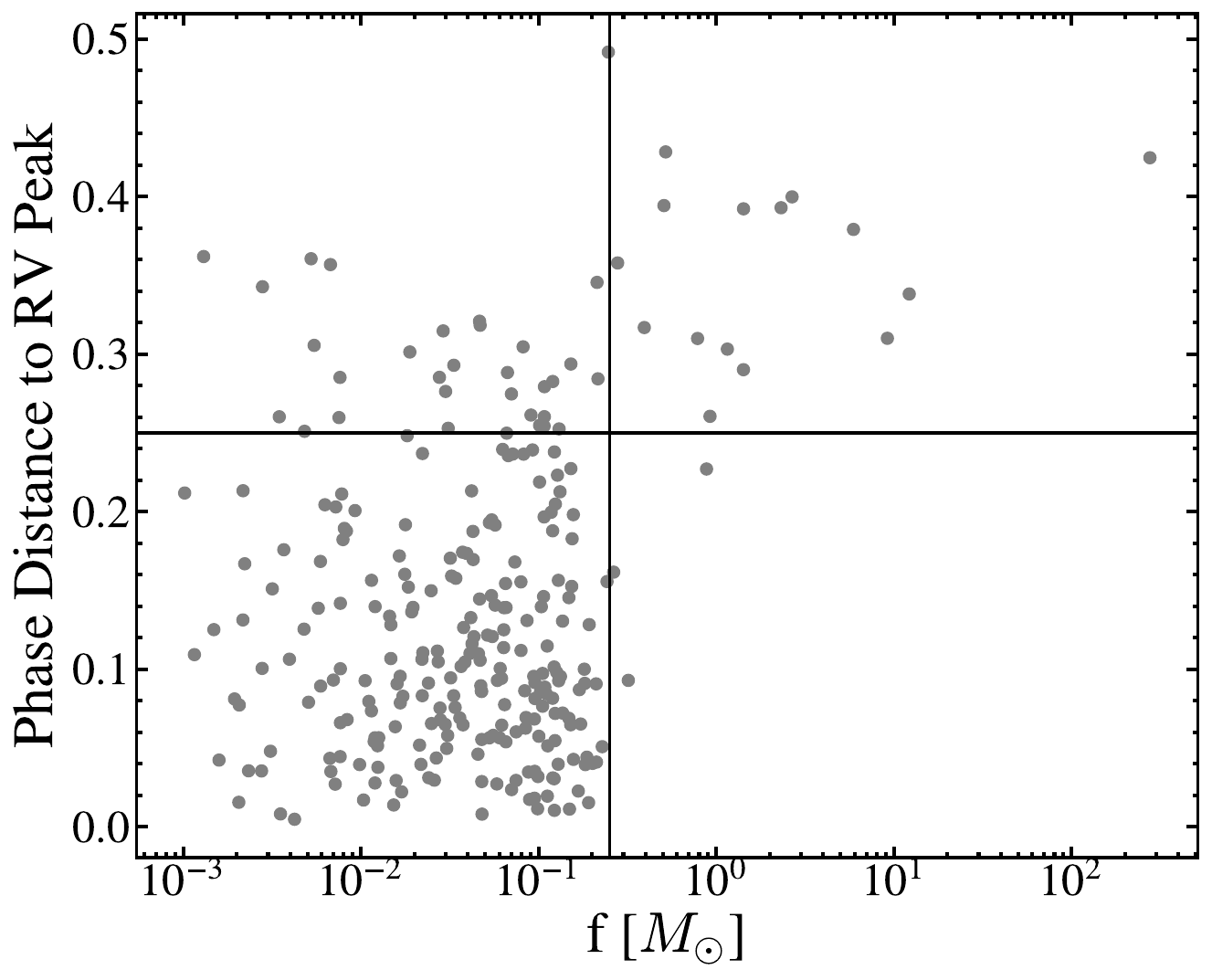}
    \caption{The maximum phase distance between an RV measurement and a velocity extrema. The vertical line at $f=0.25\ M_{\odot}$ is the mass function of an edge-on circular orbit of two $1 M_{\odot}$ stars. We discard systems above the horizontal line at a phase distance of $0.25$.}
    \label{fig:dphase_max}
\end{figure}

Some systems have RV measurements that do not well constrain both extrema of their RV curve, allowing unphysical velocity semiamplitudes and mass functions. 
To filter out these cases, we find the closest RV to the minimum and maximum velocity phases of the RV curve and the distance in phase space between these RVs and their respective peak. The larger of these two distances for each system is shown against the estimated mass function in Figure \ref{fig:dphase_max}. A vertical line at $f=0.25\ M_{\odot}$ is the mass function for a system of two $1\ M_{\odot}$ stars in an edge-on circular orbit. The systems with high mass functions all have large phase distances so we limit our sample for further analysis to exclude systems with phase distances $>0.25$. 

\begin{figure}
    \centering
    \includegraphics[width=0.97\columnwidth]{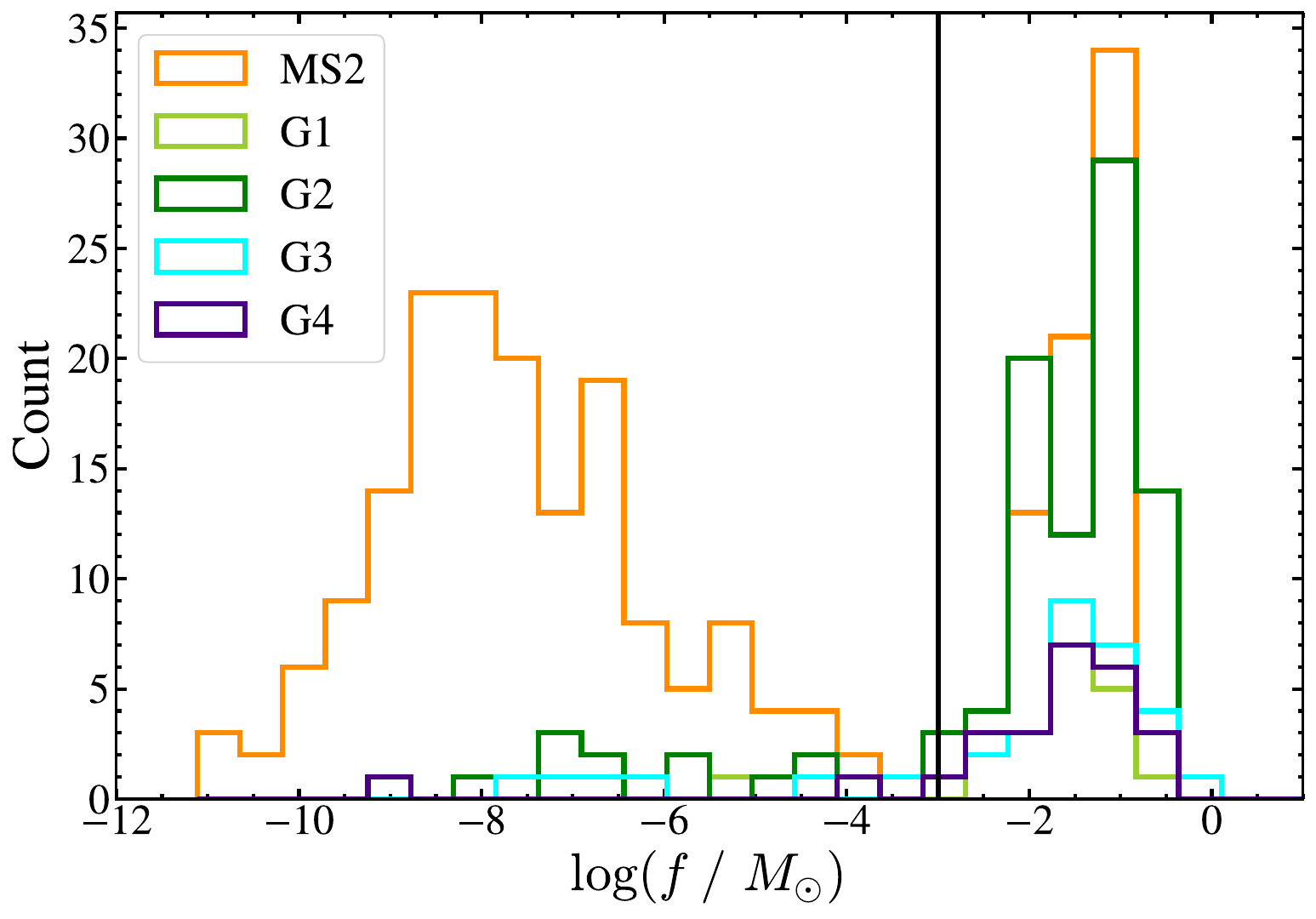}
    \caption{Distribution in binary mass function of the five groups of rotational variables from \citet{Phillips2024}. A vertical line indicates $f=0.001 M_{\odot}$, the empirical division we draw between single and binary stars.}
    \label{fig:massfunction_dist}
\end{figure}

Figure \ref{fig:massfunction_dist} shows the mass function distribution of the five groups of rotational variables. 
We empirically use $f=0.001 M_{\odot}$, denoted by a vertical line in the figure, to separate single and binary stars. Our models will correct for the tail of binaries with $f<0.001 M_{\odot}$. 
Of the 2133 APOGEE rotational variables examined in \citet{Phillips2024}, we are left with 225 systems after the criteria for the number and signal to noise ratios of the RVs, the quality of the orbital solution, and the minimum mass function are applied. There are 76, 18, 82, 26, and 23 for the MS2, G1, G2, G3, and G4 groups, respectively.

We fit the distributions in mass function using four parameters: the (average) primary mass $M_1$, the minimum mass ratio $q_{\rm{min}}$ for a uniform distribution of secondary masses $q_{\rm{min}}M_1 \leq M_2 \leq M_1$, a minimum inclination limit $i_{\rm{min}}$, and a primary mass spread $r$. 
We vary the inclination limit to model the selection effect that we do not expect to find pole-on rotators. 
The parameter $r$ allows a uniform distribution of primary masses from $M_{\rm{min}}$ to $M_{\rm{max}}$, where $M_{\rm{min}} = 2M_1/(1+r)$ and $M_{\rm{max}}=2rM_1/(1+r)$ so that the $M_1$ parameter is the average primary mass and $M_{\rm{max}}=rM_{\rm{min}}$.

\begin{figure*}
    \centering
    \includegraphics[width=\textwidth]{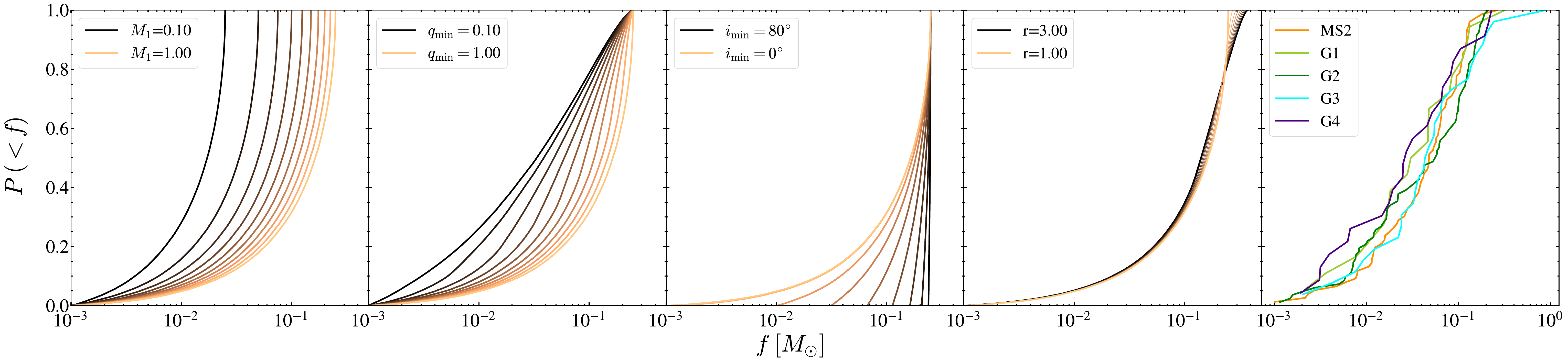}
    \caption{The integral distributions of $f$ individually varying each model parameter: the primary mass $M_1$ (leftmost), the minimum mass ratio $q_{\rm{min}}$ (center left), the inclination limit $i_{\rm{min}}$ (center), and the primary mass spread $r$ (center right). The rightmost panel shows the observed distributions in mass function for each group of rotational variables. (Leftmost panel) Distributions in $f$ assuming $r=1.0$, $q_{\rm{min}}=1.0$, and no inclination limit for 10 values of $M_1$ from 0.1--1.0$M_{\odot}$. (Center left panel) Distributions in $f$ assuming $M_1=1.0 M_{\odot}$, $r=1.0$, and no inclination limit for 10 equally spaced values of $q_{\rm{min}}$ from 0.1--1.0. (Center panel) Distributions in $f$ assuming $M_1=1.0 M_{\odot}$, $r=1$, $q_{\rm{min}}=1.0$, and inclination limits $i_{\rm{min}}=0^{\circ},10^{\circ},20^{\circ},...,80^{\circ}$. (Center right panel) Distributions in $f$ assuming $M_1 = 1.0 M_{\odot}$, $q_{\rm{min}}=1.0$ and no inclination limit for 10 equally spaced values of $r$ from 1.0--3.0.
    }
    \label{fig:param_variation}
\end{figure*}
Figure \ref{fig:param_variation} shows integral probability distributions in mass function systematically varying each of the parameters $M_1$, $q_{\rm{min}}$, $i_{\rm{min}}$, and $r$ while holding the others constant, along with the mass function distributions for each group of rotational variables. 
The average primary mass has the greatest effect on the mass function distribution, so we expect to be most sensitive to $M_1$.
The effects of varying $q_{\rm{min}}$ and $i_{\rm{min}}$ are partially degenerate, so we expect the estimates of $q_{\rm{min}}$ to be sensitive to the model's treatment of $i_{\rm{min}}$, with lower estimates of $q_{\rm{min}}$ when we have $i_{\rm{min}}\neq0$. 
Despite it's noticeable effect on the distribution, we are limited in sensitivity to low $i_{\rm{min}}$ because of our binary criteria. For example, for $M_1 = 1 M_{\odot}$ and $q=1, 0.8, 0.6, 0.4$, and $0.2$, requiring that $f>0.001 M_{\odot}$ excludes systems with $i < 9, 11, 13, 18,$ and $34$ degrees even before including any inclination limit $i_{\rm{min}}$.
The effect of varying $r$ is relatively small compared to the other parameters, so we expect limited sensitivity to the primary mass spread. Because it mainly affects the high mass function end of the distribution, we expect $r$ to also be sensitive to mass functions that have been overestimated.

It is possible that some of the mass function estimates are simply wrong either because of velocity sampling or a failure in one of our assumptions (tidally locked circular orbits). This is a particular problem at the high mass function end of the distribution where we must have $M_1 > 4f$ in order to find a non-zero probability. We accommodate this by assuming that there is some probability $p$ for a mass function value to be good, and a probability $1-p$ for it to be bad. 
We can define a vector with $v_i=0$ if $f_i$ is bad and $v_i=1$ if $f_i$ is good. The Bayesian probability of a model with $M_1$ and $q_{\rm{min}}$ is
\begin{equation}
    \label{eq:P(model|data)_unsimplified}
    P(M_1,q|D)\propto\int dp P(p) \sum_{v}\Pi_{v_i=1}pP(f_i|M_1,q)\Pi_{v_i=0}(1-p)P_B,
\end{equation}
where $P(p)$ is a prior on the probability $p$ and $P_B$ is the probability for ``bad" $f_i$ values
(\citealt{Press1997}). 
Equation \ref{eq:P(model|data)_unsimplified} can be reduced to (Kochanek, private communication to Press)
\begin{equation}
    \label{eq:P(model|data)_simplified}
    P(M_1,q|D)\propto P(M_1,q)\int dp P(p)\Pi_{i}(pP(f_i|M_1,q)+(1-p)P_B)
\end{equation}
for priors $P(M_1,q_{\rm{min}})$ on $M_1$ and $q_{\rm{min}}$. We take $P_B=1/(M_{\rm{max}}-M_{\rm{min}})$ corresponding to the range of average primary masses $M_1$ over which we perform Monte Carlo simulations (see below). 
We calculate the probability distribution of the ``good" fraction $p$ as
\begin{equation}
    \label{eq:P(p|D)}
    P(p|D)\propto P(p)\int dM_1 dq P(M_1, q)\Pi_{i}(pP(f_i|M_1,q) + (1-p)P_B).
\end{equation}

For $r=1$, we ran models with $0.1 M_{\odot} \leq M_1 \leq 2 M_{\odot}$, $0.01\leq q \leq 1$, $0\leq p < 1$ and $0^{\circ}\leq i_{\rm{min}}\leq 80^{\circ}$ with 100, 100, 100, and 9 (steps of 10 degrees) points. We also ran models with $i_{\rm{min}}=0^{\circ}$ and $i_{\rm{min}}=30^{\circ}$ using 10 values of $r$ ranging from $r=1$ to $r=3$. We did not run a full five dimensional grid of models including $r$ because, as expected, the data are not sensitive to the value of $r$. 
We assumed uniform priors for all of the variables.

To compute the mass function probability distribution for a model, we drew 100000 samples using uniform deviates for the allowed ranges of $M_1$, $M_2$, and $\cos i$, and then disposed of samples where $f<0.001 M_{\odot}$ to match our criteria for binarity. 
We sorted the remaining $N$ samples to find the integral probability distribution $P(<f_i)=i/N$. 
To estimate the differential probability of a mass function $f$ in our data, we find the index $i$ in the integral distribution where $f_i\leq f <f_{i+1}$. Using $m=500$, we find the indices $f_{i-m}$ and $f_{i+m}$ and approximate the differential probability as
\begin{equation}
    \label{eq:diff_prob_approx}
    P(f | M_1, q_{\rm{min}})=\frac{dP}{df}\simeq\frac{2m}{N}\frac{1}{f_{i+m}-f_{i-m}}
\end{equation}
(with corrections at the distribution's edges).
If $f$ lies outside the range of the simulation data, we assign the differential probability $P(f_i|M_1,q_{\rm{min}})=0$. 
To get the estimated values of any parameter, we marginalize over the
other parameters and report the median and uncertainties as the values symmetrically enclosing
90\% of the probability.  The
probability distributions for $M_1$, $q_{\rm{min}}$ and $p$ are generally
well defined so we do not show many examples of the distributions and only report the results. The distributions for $i_{\rm{min}}$ and $r$ are more problematic, and we show them to illustrate the issues and aid in the discussion.

\begin{figure*}
    \centering
    \includegraphics[width=\textwidth]{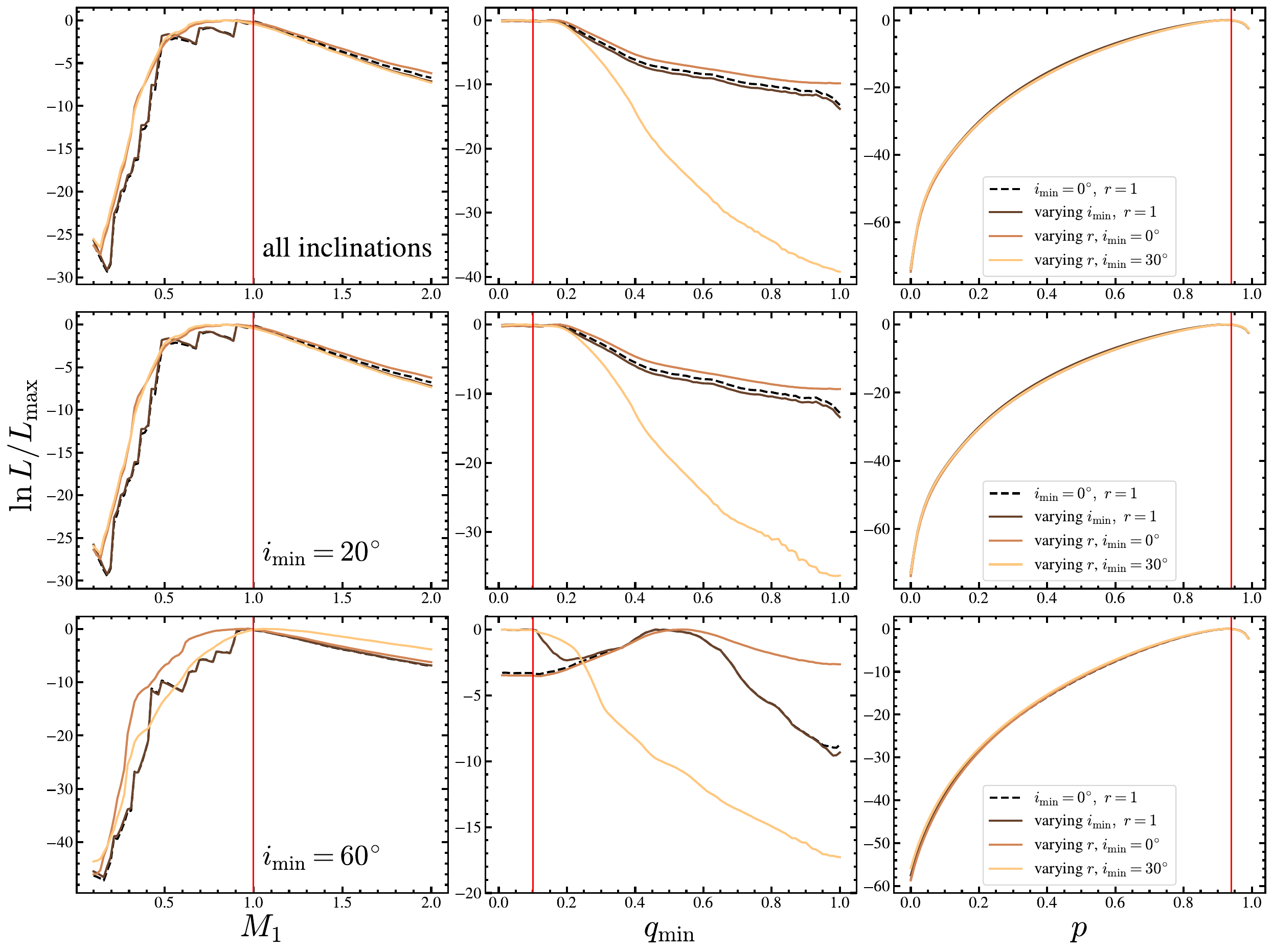}
    \caption{The likelihood of each $M_1$ (left), $q_{\rm{min}}$ (center), and $p$ (right) for the synthetic data sets allowing all inclinations (top), with $i_{\rm{min}}=20^{\circ}$ (center), and with $i_{\rm{min}}=60^{\circ}$ (bottom). The corresponding medians and 90\% confidence ranges of the integral probability for each distribution are reported in Table \ref{tab:fake_data_short}. The ground truth values of each parameter are marked with vertical red lines. Some of the distributions very closely overlap.}
    \label{fig:example_dists}
\end{figure*}

To demonstrate and test the method, we generated three synthetic data sets with 33 synthetic mass functions, the median number of systems before imposing the $f>0.001 M_{\odot}$ criteria for binarity, and with no inclination limit ($i_{\rm{min}}=0^{\circ}$), $i_{\rm{min}}=20^{\circ}$ or $i_{\rm{min}}=60^{\circ}$. We generated the synthetic data using $M_1 = 1 M_{\odot}$, $r=2.0$, $q_{\rm{min}} = 0.1$. For the data we generated, none of the mass functions were below our limit of $0.001 M_{\odot}$.
We randomly selected two of the mass functions from each data set and reset them to $1.0$ and $1.5 M_{\odot}$, meaning that the ``good" fraction for each group is $p=0.94$.
We report the resulting parameter estimates in Table \ref{tab:fake_data_short} and show the likelihood distributions for $M_1$, $q_{\rm{min}}$, and $p$ in Figure \ref{fig:example_dists}. 
The estimates of $M_1$ and $p$ are consistent with the ground truth in all cases.
As expected, there is a degeneracy between $q_{\rm{min}}$ and the inclination limit. This is particularly clear for the data set with $i_{\rm{min}}=60^{\circ}$, where, in cases where we fix $i_{\rm{min}}=0^{\circ}$, the estimate of $q_{\rm{min}}$ is inconsistent with the ground truth.
\begin{table}
    \centering
    \caption{The medians and 90\% confidence ranges for the average primary mass $M_1$ (in solar units), minimum mass ratio $q_{\rm{min}}$, and ``good" fraction $p$ for three synthetic data sets generated with $M_1 = 1M_{\odot}$, $q_{\rm{min}}=0.1$, $r=2$, $p\simeq0.94$, and either no inclination limit, $i_{\rm{min}}=20^{\circ}$, or $60^{\circ}$. We report results from each Monte Carlo fit: (1) assuming no minimum inclination and no primary mass spread, (2) marginalizing over inclination limits $i_{\rm{min}}$ but with no primary mass spread, (3) marginalizing over primary mass spread $r$ with no inclination limit ($i_{\rm{min}}=0^{\circ}$), or (4) with $i_{\rm{min}}=30^{\circ}$.}
    \begin{tabular}{r|ccc}
        \hline
        Model & $M_1\ (M_{\odot})$ & $q_{\rm{min}}$ & $p$\\
        \hline
        \multicolumn{1}{l|}{Synthetic data, $i_{\rm{min}}=0^{\circ}$} & & &\\
        \hline
        $i_{\rm{min}}=0^{\circ},\ r=1$ & $0.98 ^{+0.33}_{-0.36}$ & $0.11 ^{+0.11}_{-0.09}$ & $0.90 ^{+0.07}_{-0.11}$ \\[1mm]
        Varying $i_{\rm{min}}$, $r=1$ & $0.94 ^{+0.33}_{-0.38}$ & $0.10 ^{+0.11}_{-0.09}$ & $0.90 ^{+0.07}_{-0.12}$ \\[1mm]
        Varying $r$, $i_{\rm{min}}=0^{\circ}$ & $0.89 ^{+0.38}_{-0.27}$ & $0.12 ^{+0.12}_{-0.10}$ & $0.91 ^{+0.06}_{-0.12}$ \\[1mm]
        Varying $r$, $i_{\rm{min}}=30^{\circ}$ & $0.85 ^{+0.36}_{-0.25}$ & $0.09 ^{+0.11}_{-0.08}$ & $0.90 ^{+0.07}_{-0.11}$ \\[1mm]
       \hline
       \multicolumn{1}{l|}{Synthetic data, $i_{\rm{min}}=20^{\circ}$} & & &\\
       \hline
       $i_{\rm{min}}=0^{\circ},\ r=1$ & $0.98 ^{+0.31}_{-0.35}$ & $0.13 ^{+0.12}_{-0.11}$ & $0.90 ^{+0.07}_{-0.11}$ \\[1mm]
        Varying $i_{\rm{min}}$, $r=1$ & $0.94 ^{+0.33}_{-0.38}$ & $0.11 ^{+0.12}_{-0.09}$ & $0.90 ^{+0.07}_{-0.12}$ \\[1mm]
        Varying $r$, $i_{\rm{min}}=0^{\circ}$ & $0.89 ^{+0.38}_{-0.27}$ & $0.13 ^{+0.14}_{-0.11}$ & $0.91 ^{+0.06}_{-0.12}$ \\[1mm]
        Varying $r$, $i_{\rm{min}}=30^{\circ}$ & $0.85 ^{+0.35}_{-0.25}$ & $0.10 ^{+0.11}_{-0.09}$ & $0.90 ^{+0.07}_{-0.11}$ \\[1mm]
        \hline
        \multicolumn{1}{l|}{Synthetic data, $i_{\rm{min}}=60^{\circ}$} & & &\\
        \hline
       $i_{\rm{min}}=0^{\circ},\ r=1$ & $1.02 ^{+0.31}_{-0.12}$ & $0.49 ^{+0.14}_{-0.24}$ & $0.91 ^{+0.06}_{-0.11}$ \\[1mm]
        Varying $i_{\rm{min}}$, $r=1$ & $1.02 ^{+0.31}_{-0.12}$ & $0.41 ^{+0.20}_{-0.39}$ & $0.91 ^{+0.06}_{-0.11}$ \\[1mm]
        Varying $r$, $i_{\rm{min}}=0^{\circ}$ & $0.94 ^{+0.36}_{-0.23}$ & $0.54 ^{+0.29}_{-0.24}$ & $0.91 ^{+0.06}_{-0.11}$ \\[1mm]
        Varying $r$, $i_{\rm{min}}=30^{\circ}$ & $1.16 ^{+0.46}_{-0.25}$ & $0.09 ^{+0.12}_{-0.08}$ & $0.91 ^{+0.06}_{-0.12}$ \\[1mm]
        \hline
    \end{tabular}
    \label{tab:fake_data_short}
\end{table}

\begin{figure}
    \centering
    \includegraphics[width=\columnwidth]{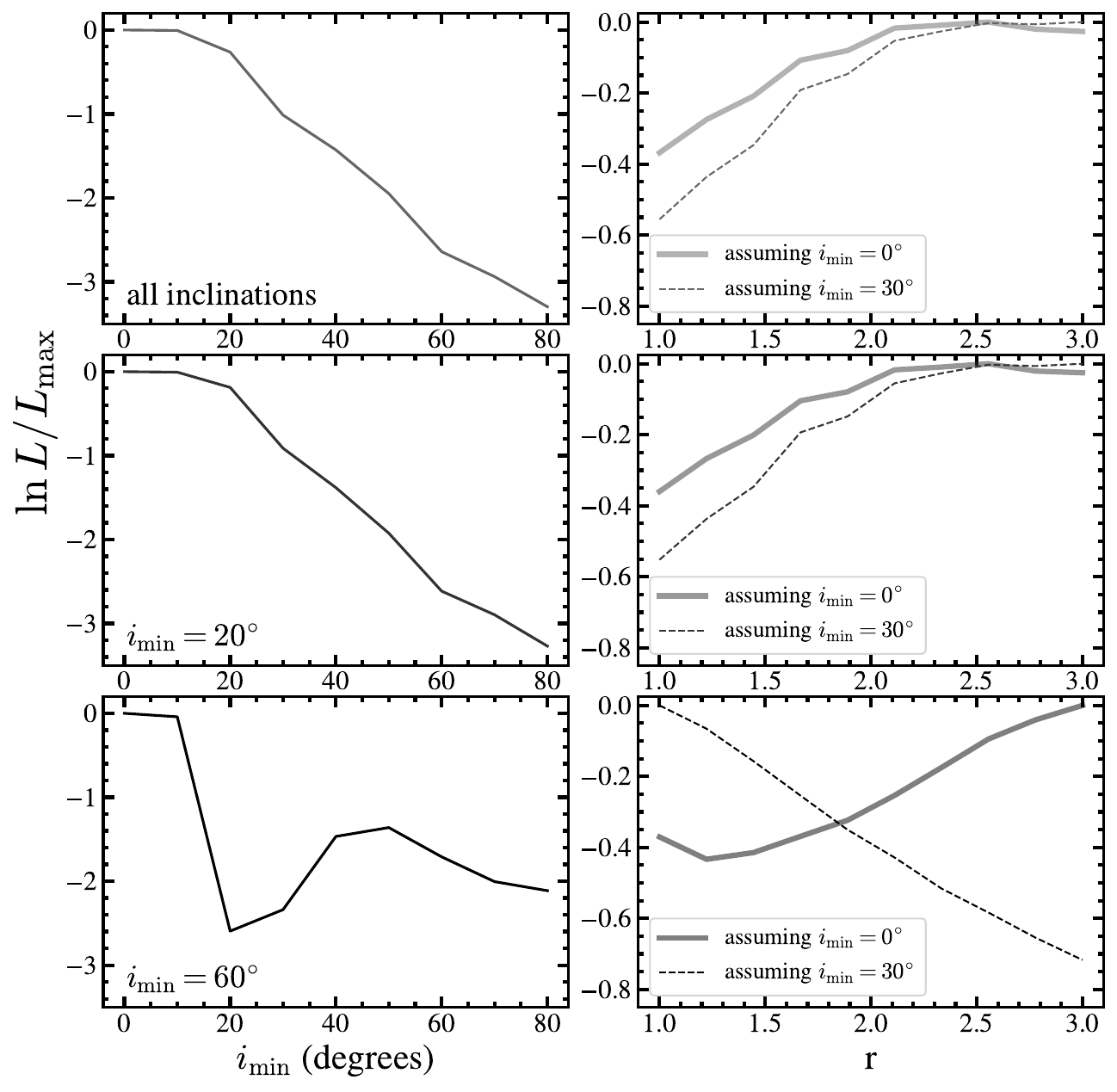}
    \caption{The likelihood of $i_{\rm{min}}$ (left) and $r$ (right) after marginalizing over  $M_1$, $q_{\rm{min}}$, and $p$ for synthetic data with $M_1 = 1\ M_{\odot},\ q_{\rm{min}}=0.1,\ r=2.0,$ and no inclination limit (top), with $i>20^{\circ}$ (center), or with $i>60^{\circ}$ (bottom). In the right panels, the solid (dashed) lines are for $i_{\rm{min}}=0^{\circ}$ ($30^{\circ}$).}
    \label{fig:L_vs_i_r_synthetic_short}
\end{figure}

The likelihoods for each inclination limit and primary mass spread for each synthetic data set are shown in Figure \ref{fig:L_vs_i_r_synthetic_short}.
All of the data sets have maximum likelihoods at $i_{\rm{min}}=0^{\circ}$, apparently favoring finding all systems with $f>0.001 M_{\odot}$. The distributions for the $i_{\rm{min}}=0^{\circ}$ and $i_{\rm{min}}=20^{\circ}$ data sets are nearly identical. 
The $i_{\rm{min}}=60^{\circ}$ data shows a local maximum at $i_{\rm{min}}=50^{\circ}$ rather that $60^{\circ}$, but does not recover the ground truth as the global likelihood maximum. 
All of the synthetic data sets have roughly flat likelihood distributions in $r$, but the likelihood differences are too small for parameter estimation.

\section{Results and Discussion}\label{sec:3}
Table \ref{tab:m*_q_values} shows the results for $M_1$, $q_{\rm{min}}$ and $p$ from our four standard fits for each group of rotational variables. 
The estimates for all three parameters are consistent between models, and each group favors little to no contamination by bad data.
The uncertainties strongly depend on the number of systems available for the analysis. They are very good for MS2 and G2 with 76 and 82 systems, and significantly worse for G1, G3 and G4, with only 18, 26, and 23 systems. 

As expected from their locations on a color magnitude diagram, the MS2 primaries are lower main sequence stars with median masses $M_1\simeq0.75 M_{\odot}$.
That the giant groups tend to have primary masses $\gtrsim1 M_{\odot}$ is required since less massive primaries have not yet evolved off of the main sequence (e.g., based on a 10 Gyr PARSEC isochrone; \citealt{parsec1, parsec2}).
The G2 subgiants favor $M_1\simeq0.94 M_{\odot}$. 
In \citet{Phillips2024}, we concluded that the G1 and G3 giants were likely a single group of RS CVn type binaries. While our estimates of $M_1$ and $q_{\rm{min}}$ for these two groups are consistent, the uncertainties are so large that this is not a strong requirement. 
Analyzed separately, the G1 group tends toward smaller primary mass estimates $M_1\simeq0.8 M_{\odot}$ (somewhat low for a red giant, but the uncertainties extend above $1 M_{\odot}$) compared to G3 at $M_1\simeq1.1 M_{\odot}$. 
Analyzing the G1 and G3 groups jointly gives $M_1\simeq 1.2 M_{\odot}$, more massive than the estimates for either of the groups on their own and the most massive of the giant groups.
The G4 group has $M_1\simeq 1.1 M_{\odot}$.

Examining the companion mass ranges, we clearly see the expected partial degeneracy with $i_{\rm{min}}$, as the models that vary $i_{\rm{min}}$ or fix $i_{\rm{min}}=30^{\circ}$ all result in lower $q_{\rm{min}}$ estimates than the models with $i_{\rm{min}}=0^{\circ}$. The MS2 stars all have a restricted range of companion masses with $q_{\rm{min}}\simeq0.4$ even though it would take $q_{\rm{min}}\simeq0.1$ to reach the Hydrogen burning limit near $0.1 M_{\odot}$ given the estimates of $M_1$. 
The giant groups all allow a broader range of companion masses with $q_{\rm{min}}\sim0.1$--$0.2$. In particular, the G4 group seems to have the weakest dependence of $q_{\rm{min}}$ on the treatment of inclination, ranging only from $q_{\rm{min}}=0.10$--$0.12$.
\begin{table}
    \centering
    \caption{The medians and 90\% confidence ranges for the average primary mass $M_1$ (in solar units), minimum mass ratio $q_{\rm{min}}$, and ``good" fraction $p$ for the MS2-G4 groups resulting from the same four fits as in Table \ref{tab:fake_data_short}.}
    \begin{tabular}{r|ccc}
    \hline
    Model & $M_1\ (M_{\odot})$ & $q_{\rm{min}}$ & $p$ \\
    \hline
    \multicolumn{1}{l|}{MS2 (76 systems)} & & &\\
    \hline
    $i_{\rm{min}}=0^{\circ},\ r=1$ & $0.75 ^{+0.12}_{-0.13}$ & $0.43 ^{+0.14}_{-0.16}$ & $0.97 ^{+0.02}_{-0.04}$ \\[1mm]
    Varying $i_{\rm{min}}$, $r=1$ & $0.75 ^{+0.12}_{-0.13}$ & $0.36 ^{+0.19}_{-0.32}$ & $0.97 ^{+0.02}_{-0.04}$ \\[1mm]
    Varying $r$, $i_{\rm{min}}=0^{\circ}$ & $0.73 ^{+0.19}_{-0.19}$ & $0.47 ^{+0.24}_{-0.19}$ & $0.98 ^{+0.01}_{-0.04}$ \\[1mm]
    Varying $r$, $i_{\rm{min}}=30^{\circ}$ & $0.83 ^{+0.21}_{-0.15}$ & $0.19 ^{+0.07}_{-0.16}$ & $0.98 ^{+0.01}_{-0.04}$ \\[1mm]
    \hline
    \multicolumn{1}{l|}{G1 (18 systems)} & & &\\
    \hline
    $i_{\rm{min}}=0^{\circ},\ r=1$ & $0.77 ^{+0.96}_{-0.21}$ & $0.22 ^{+0.28}_{-0.19}$ & $0.92 ^{+0.07}_{-0.16}$ \\[1mm]
    Varying $i_{\rm{min}}$, $r=1$ & $0.71 ^{+0.94}_{-0.17}$ & $0.15 ^{+0.27}_{-0.13}$ & $0.92 ^{+0.07}_{-0.16}$ \\[1mm]
    Varying $r$, $i_{\rm{min}}=0^{\circ}$ & $0.93 ^{+0.79}_{-0.40}$ & $0.21 ^{+0.32}_{-0.18}$ & $0.93 ^{+0.06}_{-0.16}$ \\[1mm]
    Varying $r$, $i_{\rm{min}}=30^{\circ}$ & $0.91 ^{+0.73}_{-0.36}$ & $0.13 ^{+0.16}_{-0.11}$ & $0.93 ^{+0.06}_{-0.16}$ \\[1mm]
    \hline
    \multicolumn{1}{l|}{G2 (82 systems)} & & &\\
    \hline
    $i_{\rm{min}}=0^{\circ},\ r=1$ & $0.94 ^{+0.17}_{-0.10}$ & $0.21 ^{+0.08}_{-0.16}$ & $0.99 ^{+0.00}_{-0.04}$ \\[1mm]
    Varying $i_{\rm{min}}$, $r=1$ & $0.89 ^{+0.12}_{-0.04}$ & $0.08 ^{+0.13}_{-0.06}$ & $0.99 ^{+0.00}_{-0.04}$ \\[1mm]
    Varying $r$, $i_{\rm{min}}=0^{\circ}$ & $0.96 ^{+0.19}_{-0.12}$ & $0.21 ^{+0.09}_{-0.16}$ & $0.99 ^{+0.00}_{-0.04}$ \\[1mm]
    Varying $r$, $i_{\rm{min}}=30^{\circ}$ & $0.96 ^{+0.19}_{-0.12}$ & $0.09 ^{+0.11}_{-0.07}$ & $0.99 ^{+0.00}_{-0.04}$ \\[1mm]
    \hline
    \multicolumn{1}{l|}{G3 (26 systems)} & & &\\
    \hline
    $i_{\rm{min}}=0^{\circ},\ r=1$ & $1.14 ^{+0.44}_{-0.21}$ & $0.26 ^{+0.17}_{-0.21}$ & $0.93 ^{+0.05}_{-0.13}$ \\[1mm]
    Varying $i_{\rm{min}}$, $r=1$ & $1.12 ^{+0.42}_{-0.21}$ & $0.19 ^{+0.21}_{-0.16}$ & $0.93 ^{+0.05}_{-0.13}$ \\[1mm]
    Varying $r$, $i_{\rm{min}}=0^{\circ}$ & $1.08 ^{+0.50}_{-0.31}$ & $0.28 ^{+0.19}_{-0.23}$ & $0.93 ^{+0.05}_{-0.13}$ \\[1mm]
    Varying $r$, $i_{\rm{min}}=30^{\circ}$ & $1.10 ^{+0.46}_{-0.29}$ & $0.14 ^{+0.14}_{-0.12}$ & $0.93 ^{+0.05}_{-0.13}$ \\[1mm]
    \hline
    \multicolumn{1}{l|}{G4 (23 systems)} & & &\\
    \hline
    $i_{\rm{min}}=0^{\circ},\ r=1$ & $1.12 ^{+0.50}_{-0.19}$ & $0.12 ^{+0.16}_{-0.10}$ & $0.96 ^{+0.03}_{-0.10}$ \\[1mm]
    Varying $i_{\rm{min}}$, $r=1$ & $1.10 ^{+0.48}_{-0.17}$ & $0.11 ^{+0.13}_{-0.09}$ & $0.96 ^{+0.03}_{-0.10}$ \\[1mm]
    Varying $r$, $i_{\rm{min}}=0^{\circ}$ & $1.10 ^{+0.54}_{-0.29}$ & $0.12 ^{+0.18}_{-0.10}$ & $0.96 ^{+0.03}_{-0.11}$ \\[1mm]
    Varying $r$, $i_{\rm{min}}=30^{\circ}$ & $1.06 ^{+0.48}_{-0.29}$ & $0.10 ^{+0.09}_{-0.09}$ & $0.96 ^{+0.03}_{-0.11}$ \\[1mm]
    \hline
    \multicolumn{1}{l|}{G1+G3 (44 systems)} &&&\\
    \hline
    $i_{\rm{min}}=0^{\circ},\ r=1$ & $1.29 ^{+0.31}_{-0.42}$ & $0.20 ^{+0.15}_{-0.17}$ & $0.94 ^{+0.05}_{-0.09}$ \\[1mm]
    Varying $i_{\rm{min}}$, $r=1$ & $1.21 ^{+0.36}_{-0.36}$ & $0.16 ^{+0.16}_{-0.14}$ & $0.94 ^{+0.05}_{-0.09}$ \\[1mm]
    Varying $r$, $i_{\rm{min}}=0^{\circ}$ & $1.08 ^{+0.42}_{-0.29}$ & $0.24 ^{+0.16}_{-0.19}$ & $0.95 ^{+0.04}_{-0.09}$ \\[1mm]
    Varying $r$, $i_{\rm{min}}=30^{\circ}$ & $1.06 ^{+0.38}_{-0.25}$ & $0.13 ^{+0.10}_{-0.11}$ & $0.95 ^{+0.04}_{-0.09}$ \\[1mm]
    \hline
    \end{tabular}
    \label{tab:m*_q_values}
\end{table}

Since more luminous main sequence primaries should have higher masses, we split the MS2 group of rotators into three bins in \textit{Gaia} absolute G magnitude with equal numbers of objects per magnitude bin. The estimates for $M_1$, $q_{\rm{min}}$, and $p$ for each magnitude bin are reported in Table \ref{tab:ms2split_mstar_q_values}. Each bin favors little contamination by bad data.
We also report mass ranges for a 10 Gyr PARSEC isochrone \citep{parsec1,parsec2} for each magnitude bin after doubling the luminosity at fixed color to align it with the ``binary" main sequence. As expected, the primary masses increase with increasing luminosity, with estimates of $\sim0.62 M_{\odot}$ in the dimmest bin and $\sim0.90 M_{\odot}$ in the brightest bin. The 90\% confidence regions for primary mass estimates are consistent with the isochrone mass ranges in each bin. 

There is a selection effect pointed out by \cite{simonian2019} that can affect estimates of $q_{\rm{min}}$ for the
MS2 sample.  If luminosity on the main sequence scales as $L\propto M^3$, then the luminosity of a binary is $L\propto (1+q^3)$, so the distance at which a binary can be found in a flux-limited sample scales as
$(1+q^3)^{1/2}$.  For a constant density, spherical distribution, the number of binaries would then
scale as $(1+q^3)^{3/2}$ --- the actual scaling will be weaker because most of the stars are in a disk, which
asymptotically reduces the scaling to $(1+q^2)$, and dust. Such a selection effect reduces the numbers
of lower $q$ systems in the sample, and thus would bias the models to higher values of $q_{\rm{min}}$.  We
do not explore this in detail here because doing it correctly is a significant project that seems unwarranted
given our small sample size, which make the statistical uncertainties on $q_{\rm{min}}$ significant, and because the parallax error requirement on the sample
(see \citealt{Phillips2024}) that $\varpi/\sigma_\varpi>10$ 
also makes much of our sample volume limited rather than flux limited.  In a volume limited
sample there is no such bias on $q_{\rm{min}}$.  This question will need to be addressed for larger samples.

\begin{table}
    \centering
    \caption{The medians and 90\% confidence ranges for the average primary mass $M_1$, minimum mass ratio $q_{\rm{min}}$, and ``good" fraction $p$ for the MS2 group split into three bins in \textit{Gaia} absolute $G$ magnitude for the same four fits as in Tables \ref{tab:fake_data_short} and \ref{tab:m*_q_values}. We also report the mass ranges for a 10 Gyr binary PARSEC isochrone in each magnitude bin, $M_{\rm{iso}}$.}
    \begin{tabular}{r|ccc}
        \hline
        Model & $M_1\ (M_{\odot})$ & $q_{\rm{min}}$ & $p$ \\
        \hline
        \multicolumn{1}{l|}{$4.78 \leq M_G < 5.91$}&&\\
        \multicolumn{1}{l|}{$M_{\rm{iso}}=0.71$--$0.82\ M_{\odot}$}&&\\
        \hline
        $i_{\rm{min}}=0^{\circ},\ r=1$ & $0.81 ^{+0.63}_{-0.19}$ & $0.43 ^{+0.36}_{-0.38}$ & $0.95 ^{+0.04}_{-0.12}$ \\[1mm]
        Varying $i_{\rm{min}}$, $r=1$ & $0.81 ^{+0.61}_{-0.19}$ & $0.13 ^{+0.53}_{-0.11}$ & $0.95 ^{+0.04}_{-0.12}$ \\[1mm]
        Varying $r$, $i_{\rm{min}}=0^{\circ}$ & $0.91 ^{+0.50}_{-0.38}$ & $0.44 ^{+0.42}_{-0.37}$ & $0.96 ^{+0.03}_{-0.11}$ \\[1mm]
        Varying $r$, $i_{\rm{min}}=30^{\circ}$ & $1.08 ^{+0.46}_{-0.33}$ & $0.09 ^{+0.12}_{-0.08}$ & $0.96 ^{+0.03}_{-0.10}$ \\[1mm]
        \hline
        \multicolumn{1}{l|}{$5.92 \leq M_G < 6.54$}&&\\
        \multicolumn{1}{l|}{$M_{\rm{iso}}=0.65$--$0.71\ M_{\odot}$} &&\\
        \hline
        $i_{\rm{min}}=0^{\circ},\ r=1$ & $0.81 ^{+0.31}_{-0.21}$ & $0.40 ^{+0.20}_{-0.20}$ & $0.96 ^{+0.03}_{-0.09}$ \\[1mm]
        Varying $i_{\rm{min}}$, $r=1$ & $0.81 ^{+0.29}_{-0.21}$ & $0.33 ^{+0.23}_{-0.28}$ & $0.96 ^{+0.03}_{-0.10}$ \\[1mm]
        Varying $r$, $i_{\rm{min}}=0^{\circ}$ & $0.73 ^{+0.35}_{-0.21}$ & $0.46 ^{+0.30}_{-0.22}$ & $0.96 ^{+0.03}_{-0.09}$ \\[1mm]
        Varying $r$, $i_{\rm{min}}=30^{\circ}$ & $0.81 ^{+0.36}_{-0.21}$ & $0.26 ^{+0.15}_{-0.21}$ & $0.96 ^{+0.03}_{-0.09}$ \\[1mm]
        \hline
        \multicolumn{1}{l|}{$6.55 \leq M_G < 8.57$}&&\\
        \multicolumn{1}{l|}{$M_{\rm{iso}}=0.47$--$0.65\ M_{\odot}$}&&\\
        \hline
       $i_{\rm{min}}=0^{\circ},\ r=1$ & $0.62 ^{+0.25}_{-0.10}$ & $0.35 ^{+0.30}_{-0.23}$ & $0.97 ^{+0.02}_{-0.09}$ \\[1mm]
        Varying $i_{\rm{min}}$, $r=1$ & $0.60 ^{+0.25}_{-0.08}$ & $0.23 ^{+0.32}_{-0.20}$ & $0.97 ^{+0.02}_{-0.09}$ \\[1mm]
        Varying $r$, $i_{\rm{min}}=0^{\circ}$ & $0.60 ^{+0.31}_{-0.17}$ & $0.39 ^{+0.45}_{-0.25}$ & $0.97 ^{+0.02}_{-0.09}$ \\[1mm]
        Varying $r$, $i_{\rm{min}}=30^{\circ}$ & $0.64 ^{+0.31}_{-0.13}$ & $0.24 ^{+0.14}_{-0.20}$ & $0.97 ^{+0.02}_{-0.09}$ \\[1mm]
        \hline
    \end{tabular}
    \label{tab:ms2split_mstar_q_values}
\end{table}

\begin{figure*}
    \centering
    \includegraphics[width=0.75\textwidth]{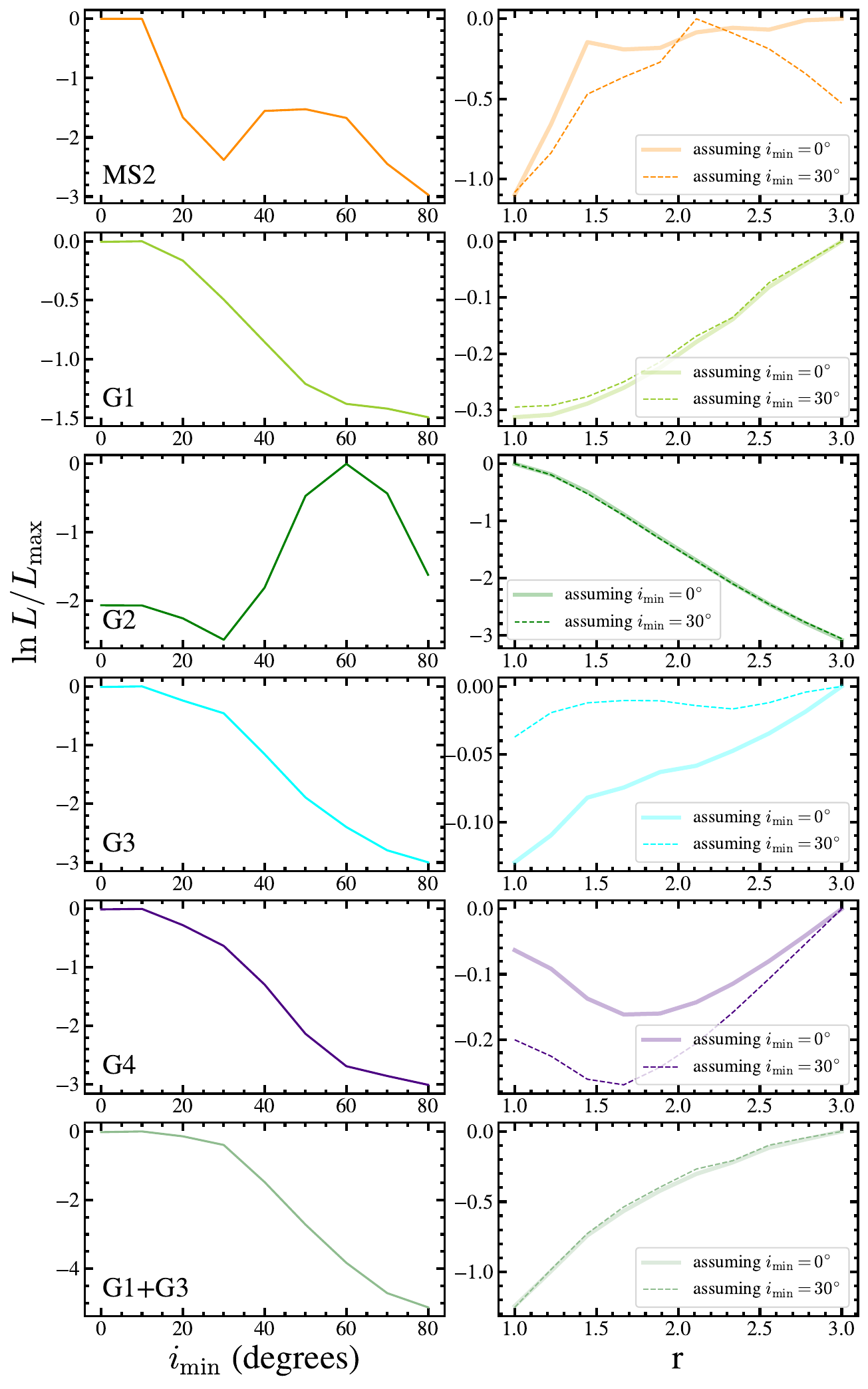}
    \caption{The likelihood of $i_{\rm{min}}$ (left) and $r$ (right) after marginalizing over $p$, $M_1$, and $q_{\rm{min}}$ for each group of rotational variables. In the right panels, the solid (dashed) lines are for $i_{\rm{min}}=0^{\circ}$ ($30^{\circ}$).}
    \label{fig:L_vs_i_r}
\end{figure*}

Figure \ref{fig:L_vs_i_r} shows the likelihood distributions for the inclination limit and primary mass spread (both assuming $i_{\rm{min}}=0^{\circ}$ and $i_{\rm{min}}=30^{\circ}$). We discuss the structure of likelihood differences across $i_{\rm{min}}$ and $r$, but note that the differences are not particularly significant.
With the exception of the G2 group, there is no strong indication of an inclination bias (up to the insensitivity we discussed in Section \ref{sec:2}). G1, G3 (both separated and combined), and G4 favor finding most if not all systems with $f>0.001 M_{\odot}$ regardless of inclination, with maximum likelihoods $i_{\rm{min}}=10^{\circ}$.
MS2 has a maximum likelihood at $i_{\rm{min}}=0^{\circ}$, but its likelihood does not monotonically decrease with increasing $i_{\rm{min}}$. Instead, it mirrors the behavior of the synthetic data in the bottom left panel of Figure \ref{fig:L_vs_i_r_synthetic_short}, with a local maximum at $i_{\rm{min}}=50^{\circ}$, suggesting a potential inclination limit $\gtrsim30^{\circ}$. 
The G2 group has a maximum likelihood at $i_{\rm{min}}=60^{\circ}$, preferring to be viewed quasi-equatorially.

For the values of $r$, all but the G2 group have roughly flat distributions, with little distinction between assuming $i_{\rm{min}}=0^{\circ}$ and $i_{\rm{min}}=30^{\circ}$. The MS2 group has maximum likelihoods at $r=3.0$ and $r=2.11$ when we assume $i_{\rm{min}}=0^{\circ}$ and $i_{\rm{min}}=30^{\circ}$, respectively. 
The G1, G3 (both separated and combined), and G4 groups have maximum likelihoods at $r=3$ regardless of whether we assume $i_{\rm{min}}=0^{\circ}$ or $30^{\circ}$. 
The G2 group favors no spread in primary mass, with a maximum likelihood at $r=1$. 

\section{Conclusions}\label{sec:4}

We presented a method to estimate the average primary mass, secondary mass range, inclination limit, primary mass spread, and fraction of correctly estimated binary mass functions for populations binary stars, and applied it to five previously identified populations of binary rotational variables. 
Given that we recover consistent values of the good data probability $p$ with the ground truth in all cases of synthetic data, we are reasonably confident that we correctly estimate it. Each group of rotational variables favors little to no contamination by incorrectly estimated mass functions.
    
We are also reasonably confident in our estimates of average primary mass, which we also recover correctly in the synthetic data tests. 
In the rotational variables, we find MS2 stars to have average primary masses of $\sim 0.75 M_{\odot}$, as expected from their location on the CMD.
We estimate different average primary masses for G1 ($\sim0.8 M_{\odot}$) and G3 ($\sim1.10 M_{\odot}$) when analyzing them separately, and higher masses ($\sim1.2 M_{\odot}$) when analyzing them jointly. The uncertainties on the separated G1 and G3 mass estimates are large enough that we are still unable to definitively say whether they are physically different groups, or a single group as we posited in \citet{Phillips2024}. Regardless, all of these mass estimates are consistent with the $1$--$2 M_{\odot}$ upper limit on RS CVn primaries from \citet{Leiner2022}. The G2 stars are somewhat less massive than the G1 and G3 groups analyzed jointly, at $\sim 0.94 M_{\odot}$, consistent with their characterization as sub-subgiants which should be less massive than typical RS CVn. For the G4 group we estimate average primary masses intermediate to those in G1-3, around $1.10 M_{\odot}$. We also confirm that our estimates of $M_1$ increase with increasing luminosity for the MS2 main sequence binaries.

For populations without a strong inclination bias, we are also relatively confident in our estimates of minimum mass ratio. We correctly recover $q_{\rm{min}}$ in most cases of synthetic data, failing to do so only when we assume $i_{\rm{min}}=0^{\circ}$ for synthetic data constructed with a high inclination limit. The estimates of $q_{\rm{min}}$ are partially degenerate with the model's treatment of inclination, and this effect becomes stronger at higher true inclination limits in synthetic data. The MS2 stars seem to have a restricted range of companion masses, with estimates of $q_{\rm{min}}\sim0.4$, though we have reason to believe they have an inclination limit above our insensitivity threshold of $\sim20^{\circ}$ (see below) and therefore may actually have companions down to lower mass ratios. The giant groups have $q_{\rm{min}}\simeq0.1$--$0.2$, with the weakest dependence of $q_{\rm{min}}$ on $i_{\rm{min}}$ for the G4 group.
    
We are not sensitive to $i_{\rm{min}}\lesssim 20$--$30^{\circ}$ because of our minimum mass function criteria (excluding $f<0.001$). The G1 and G3 (analyzed both separately and jointly) and G4 stars all show maximum likelihood inclination limits near $0^{\circ}$, favoring finding all systems with $f>0.001 M_{\odot}$ and indicating no strong inclination bias. For higher inclination limits, our tests with synthetic data show that we are only somewhat sensitive to $i_{\rm{min}}$, observing local maxima in likelihood at high inclinations, but with a maximum likelihood at $i_{\rm{min}}=0^{\circ}$. Such a shape of the $i_{\rm{min}}$ likelihood distribution may be useful as a diagnostic for whether it is necessary to consider parameter estimates from models varying $i_{\rm{min}}$. This is the case for the MS2 group which has a local maximum at $i_{\rm{min}}=40^{\circ}$, much like the synthetic data with $i_{\rm{min}}=60^{\circ}$, indicating that it may have a true inclination limit $\gtrsim 30$--$50^{\circ}$. 

We are not sensitive to the primary mass spread and are unable to recover the correct value for $r$ as the maximum likelihood in the synthetic data. The likelihoods in $r$ are relatively flat both for the synthetic data and for each group of rotational variables, providing no evidence for a preferred primary mass spread. 

The volume of data available for this type of analysis should increase rapidly increase with the SDSS Milky Way Mapper \citep{MWM2017} extension of the APOGEE survey and \textit{Gaia} DR4. We explored the expected improvement by repeating each of the synthetic mass function tests with 10 times as much data (including 10 times as many ``bad" mass functions).
We report the results for $M_1$, $q_{\rm{min}}$, and $p$ from each model in Table \ref{tab:fake_data_long}. 
The primary mass estimates are consistent with the ground truth value of $1 M_{\rm{\odot}}$, are more uniform across models, and have significantly smaller uncertainties compared to those in Table \ref{tab:fake_data_short}, especially for the data with $i_{\rm{min}}=0^{\circ}$ and $20^{\circ}$. 
In the minimum mass fractions, we still see the dependence of $q_{\rm{min}}$ on inclination. For the $i_{\rm{min}}=0^{\circ}$ and $i_{\rm{min}}=20^{\circ}$ data sets, even the values of $q_{\rm{min}}$ where the model assumes no inclination limit are consistent with the ground truth value of $q_{\rm{min}}=0.1$. For the $i_{\rm{min}}=60^{\circ}$ data set, the values of $q_{\rm{min}}$ when the model fixes $i_{\rm{min}}=0^{\circ}$ ($q_{\rm{min}}\sim0.46$) are inconsistent with the ground truth, but given that we will be better able to constrain the inclination limits with more data (see below), this is less of a cause for concern.
\begin{table}
    \centering
    \label{tab:fake_data_long}
    \caption{The estimates and 90\% confidence ranges for the average primary mass $M_1$ (in solar units), minimum mass ratio $q_{\rm{min}}$, and ``good" fraction $p$ for three 10 times larger synthetic data sets generated with $M_1=1 M_{\odot},\ q_{\rm{min}}=0.1,\ r=2$, and either $i_{\rm{min}}=0^{\circ}$, $i_{\rm{min}}=20^{\circ}$, or $i_{\rm{min}}=60^{\circ}$ for the same 4 fits as in Tables \ref{tab:fake_data_short}--\ref{tab:ms2split_mstar_q_values}.}
    \begin{tabular}{r|ccc}
        \hline
        Model & $M_1$ & $q_{\rm{min}}$ & $p$ \\
        \hline
        \multicolumn{1}{l|}{Synthetic data, $i_{\rm{min}}=0$}&&&\\
        \hline
        $i_{\rm{min}}=0^{\circ},\ r=1$ & $0.96 ^{+0.08}_{-0.06}$ & $0.11 ^{+0.06}_{-0.09}$ & $0.92 ^{+0.03}_{-0.03}$ \\[1mm]
        Varying $i_{\rm{min}}$, $r=1$ & $0.96 ^{+0.08}_{-0.06}$ & $0.10 ^{+0.06}_{-0.08}$ & $0.92 ^{+0.03}_{-0.03}$ \\[1mm]
        Varying $r$, $i_{\rm{min}}=0^{\circ}$ & $0.96 ^{+0.08}_{-0.08}$ & $0.11 ^{+0.07}_{-0.09}$ & $0.92 ^{+0.03}_{-0.03}$ \\[1mm]
        Varying $r$, $i_{\rm{min}}=30^{\circ}$ & $0.94 ^{+0.08}_{-0.10}$ & $0.08 ^{+0.05}_{-0.06}$ & $0.92 ^{+0.03}_{-0.03}$ \\[1mm]
        \hline
        \multicolumn{1}{l|}{Synthetic data, $i_{\rm{min}}=20^{\circ}$}&&&\\
        \hline
        $i_{\rm{min}}=0^{\circ},\ r=1$ & $1.02 ^{+0.06}_{-0.10}$ & $0.13 ^{+0.06}_{-0.11}$ & $0.92 ^{+0.03}_{-0.03}$ \\[1mm]
        Varying $i_{\rm{min}}$, $r=1$ & $1.00 ^{+0.06}_{-0.10}$ & $0.11 ^{+0.07}_{-0.09}$ & $0.92 ^{+0.03}_{-0.03}$ \\[1mm]
        Varying $r$, $i_{\rm{min}}=0^{\circ}$ & $0.98 ^{+0.08}_{-0.08}$ & $0.14 ^{+0.05}_{-0.11}$ & $0.92 ^{+0.03}_{-0.03}$ \\[1mm]
        Varying $r$, $i_{\rm{min}}=30^{\circ}$ & $0.96 ^{+0.08}_{-0.10}$ & $0.08 ^{+0.06}_{-0.06}$ & $0.92 ^{+0.03}_{-0.03}$ \\[1mm]
        \hline
        \multicolumn{1}{l|}{Synthetic data, $i_{\rm{min}}=60^{\circ}$}&&&\\
        \hline
        $i_{\rm{min}}=0^{\circ},\ r=1$ & $1.10 ^{+0.06}_{-0.06}$ & $0.32 ^{+0.08}_{-0.06}$ & $0.92 ^{+0.03}_{-0.02}$ \\[1mm]
        Varying $i_{\rm{min}}$, $r=1$ & $1.10 ^{+0.06}_{-0.08}$ & $0.07 ^{+0.05}_{-0.06}$ & $0.92 ^{+0.03}_{-0.02}$ \\[1mm]
        Varying $r$, $i_{\rm{min}}=0^{\circ}$ & $1.08 ^{+0.06}_{-0.10}$ & $0.33 ^{+0.11}_{-0.06}$ & $0.92 ^{+0.03}_{-0.02}$ \\[1mm]
        Varying $r$, $i_{\rm{min}}=30^{\circ}$ & $1.12 ^{+0.08}_{-0.06}$ & $0.08 ^{+0.04}_{-0.07}$ & $0.92 ^{+0.03}_{-0.02}$ \\[1mm]
        \hline
    \end{tabular}
\end{table}

\begin{figure}
    \centering
    \includegraphics[width=\columnwidth]{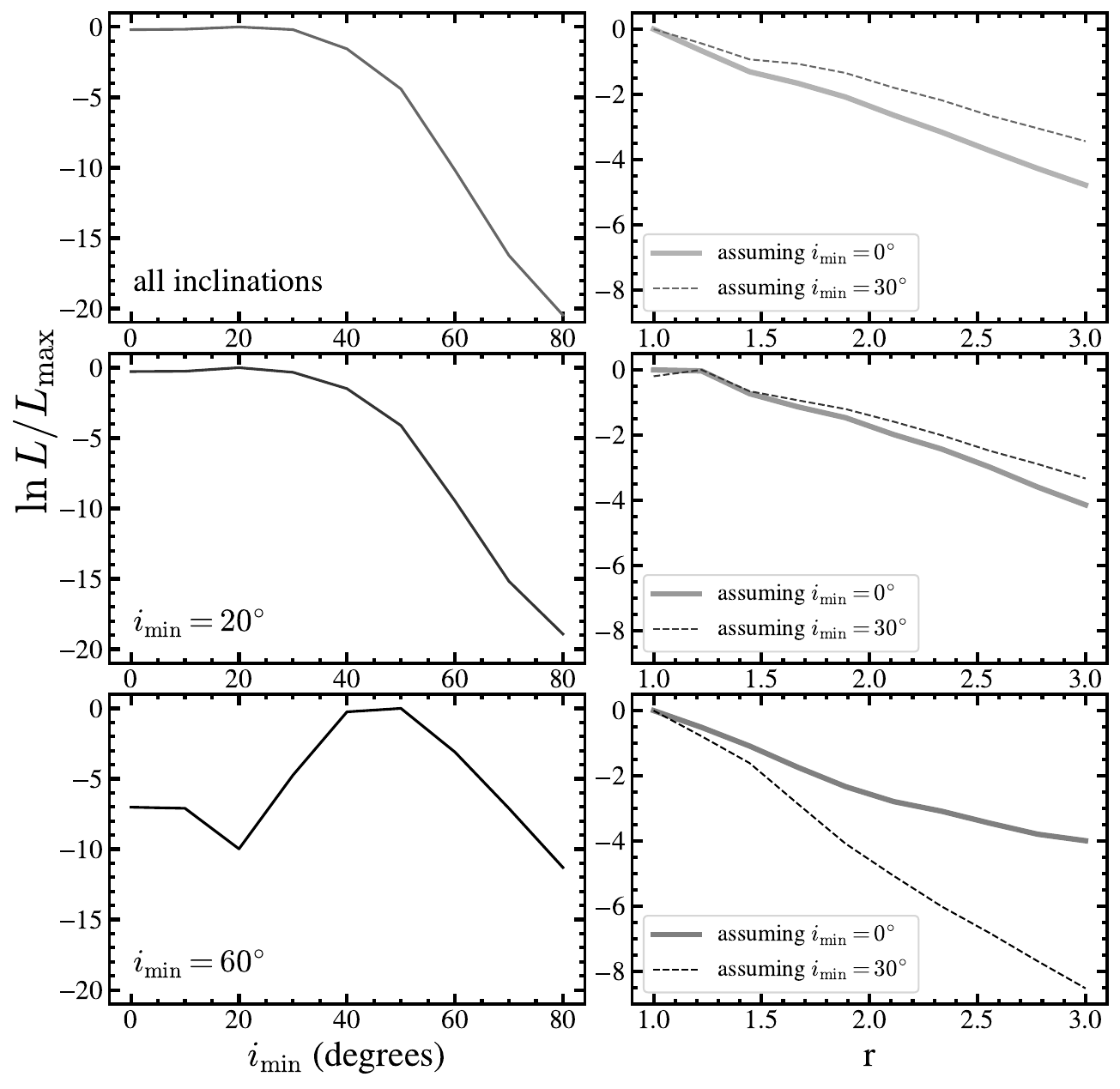}
    \caption{The likelihood of $i_{\rm{min}}$ (left) and $r$ (right) after marginalizing over $p$, $M_1$, and $q_{\rm{min}}$ for a larger synthetic data set compared to the one described in Section \ref{sec:2} with $M_1 = 1\ M_{\odot},\ q_{\rm{min}}=0.1,\ r=2.0,$ and no inclination limit (top), with $i>20^{\circ}$ (center), or with $i>60^{\circ}$ (bottom). In the right panels, the solid (dashed) lines are for $i_{\rm{min}}=0^{\circ}$ ($30^{\circ}$).}
    \label{fig:L_vs_i_r_synthetic_long}
\end{figure}

Figure \ref{fig:L_vs_i_r_synthetic_long} shows the likelihoods of $i_{\rm{min}}$ and $r$ for the fits to the longer synthetic data tests. 
For the inclination limits, the data with no inclination limit and with $i_{\rm{min}}=20^{\circ}$ both have maximum likelihoods at at $i_{\rm{min}}=20^{\circ}$, and decrease steeply in likelihood at higher inclination limits. The $i_{\rm{min}}=60^{\circ}$ data set has a maximum likelihood at $i_{\rm{min}}=50^{\circ}$, meaning that with $\sim10$ times as much data, we will be better able to constrain $i_{\rm{min}}$.

The likelihoods for the primary mass spreads are still flat compared to the likelihood distributions for $i_{\rm{min}}$. 
All fits favor no primary mass spread, with maximum likelihoods at or near $r=1$, though the $i_{\rm{min}}=0$ and $i_{\rm{min}}=20$ data sets show hints at local maxima near $r=2$.
We checked whether the inclusion of ``bad" mass functions was affecting our estimates of $r$ by generating another synthetic data set with no bad points and looking at the results with $p\equiv 1$. This yielded nearly identical results. Thus, even with ten times the data it will be difficult to constrain $r$.

These tests show that with increasingly large samples, the constraining power of this approach increases significantly, both in terms of better constraints on the variables that were reasonably well-constrained for the present sample ($M_1$ and $q_{\rm{min}}$) but also will begin to constrain the variables that were not well-constrained ($i_{\rm{min}}$ and $r$). SDSS Milky Way Mapper \citep{MWM2017} is likely to provide an order of magnitude increase, and \textit{Gaia} DR4 is likely to provide an even larger expansion.  The ASAS-SN sample of rotational variables has a limited magnitude overlap with good \textit{Gaia} spectroscopic observations, so expanding the rotational variable sample being used to brighter magnitudes (e.g., the samples from \citealt{oelkers2018}, \citealt{chen2020}, \citealt{reinhold2023}, or \citealt{claytor2024}) can provide a still larger increase.  The method can also be applied to any population of SB1 binaries providing only mass functions.

\section*{Acknowledgements}

The authors thank Las Cumbres Observatory and its staff for their
continued support of ASAS-SN. 
AP thanks Dominick Rowan and Dr. Christine Mazzola Daher for their guidance in using and interpreting APOGEE RV data, and Profs. Krzysztof Stanek and Marc Pinsonneault for their helpful comments in preparing the manuscript. 
CSK is supported by National Science Foundation
grants AST-2307385 and 2407206.

ASAS-SN is funded in part by the Gordon and Betty Moore
Foundation through grants GBMF5490 and GBMF10501, and by the
Alfred P. Sloan Foundation through grant G-2021-14192
to the Ohio State University.


Funding for the Sloan Digital Sky 
Survey IV has been provided by the 
Alfred P. Sloan Foundation, the U.S. 
Department of Energy Office of 
Science, and the Participating 
Institutions. 

SDSS-IV acknowledges support and 
resources from the Center for High 
Performance Computing  at the 
University of Utah. The SDSS 
website is www.sdss4.org.

SDSS-IV is managed by the 
Astrophysical Research Consortium 
for the Participating Institutions 
of the SDSS Collaboration including 
the Brazilian Participation Group, 
the Carnegie Institution for Science, 
Carnegie Mellon University, Center for 
Astrophysics | Harvard \& 
Smithsonian, the Chilean Participation 
Group, the French Participation Group, 
Instituto de Astrof\'isica de 
Canarias, The Johns Hopkins 
University, Kavli Institute for the 
Physics and Mathematics of the 
Universe (IPMU) / University of 
Tokyo, the Korean Participation Group, 
Lawrence Berkeley National Laboratory, 
Leibniz Institut f\"ur Astrophysik 
Potsdam (AIP),  Max-Planck-Institut 
f\"ur Astronomie (MPIA Heidelberg), 
Max-Planck-Institut f\"ur 
Astrophysik (MPA Garching), 
Max-Planck-Institut f\"ur 
Extraterrestrische Physik (MPE), 
National Astronomical Observatories of 
China, New Mexico State University, 
New York University, University of 
Notre Dame, Observat\'ario 
Nacional / MCTI, The Ohio State 
University, Pennsylvania State 
University, Shanghai 
Astronomical Observatory, United 
Kingdom Participation Group, 
Universidad Nacional Aut\'onoma 
de M\'exico, University of Arizona, 
University of Colorado Boulder, 
University of Oxford, University of 
Portsmouth, University of Utah, 
University of Virginia, University 
of Washington, University of 
Wisconsin, Vanderbilt University, 
and Yale University.


\section*{Data Availability}
All data used in this study are public. 
The catalogue of ASAS-SN variable stars is available on the ASAS-SN variable stars
database (\url{https://asas-sn.osu.edu/variables}). 
APOGEE DR17 radial velocities are available in the \texttt{allvisit} file at \url{https://www.sdss4.org/dr17/irspec/spectro_data/}.



\bibliographystyle{mnras}
\bibliography{ref}







\bsp	
\label{lastpage}
\end{document}